\documentclass
[
amssymb,prl,aps,twocolumn,floatfix,tightenlines,superscriptaddress]{revtex4}
\usepackage{graphicx}
\usepackage{color}
\usepackage{eurosym}
\usepackage{dcolumn}
\usepackage{bm}
\usepackage{subfigure}
\usepackage[final]{pdfpages}
\usepackage{setspace}
\usepackage{pdfpages}
\usepackage{amsmath}

\begin{document}
\title{Complementary metal-oxide semiconductor compatible source of single photons at near-visible wavelengths}

\author{Robert Cernansky}
\affiliation{School of Physics and Astronomy, University of Southampton, Southampton, SO17 1BJ, United Kingdom}
\author{Francesco Martini}
\affiliation{School of Physics and Astronomy, University of Southampton, Southampton, SO17 1BJ, United Kingdom}
\author{Alberto Politi}
\affiliation{School of Physics and Astronomy, University of Southampton, Southampton, SO17 1BJ, United Kingdom}
\email{A.Politi@soton.ac.uk}

\begin{abstract}
We demonstrate on chip generation of correlated pairs of photons in the near-visible spectrum using a CMOS compatible PECVD Silicon Nitride photonic device. Photons are generated via spontaneous four wave mixing enhanced by a ring resonator with high quality Q-factor of 320,000 resulting in a generation rate of 950,000 $\frac{pairs}{mW}$. The high brightness of this source offers the opportunity to expand photonic quantum technologies over a broad wavelength range and provides a path to develop fully integrated quantum chips working at room temperature.
\end{abstract}

\maketitle

Non-classical states of light generated through spontaneous four wave mixing (SFWM) has been an essential source for integrated quantum photonic structures \cite{Sharping2006, Takesue2007a, Clemmen2009}. Highly enhanced light-matter interactions can be achieved by microring resonators due to constructing interference yielding to an efficient production of time-bin entangled photons \cite{Azzini2012, Grassani2015} useful for quantum computation \cite{Knill2001}, simulation \cite{Aspuru-Guzik2012} and quantum key distribution \cite{Gisin2007}. The enhanced generation rate of single photons relies on high quality factor $Q$ and small modal volume $V$ of the resonator, therefore low losses and small rings are highly desirable.
So far there has been an intensive investigation of on chip integrated sources generating photons at telecom wavelengths \cite{Silverstone2013, Jin2014b, Preble2015}. This focus has been highly motivated not only because of the advanced technologies in integrated silicon photonics \cite{Xiong2016} and low loss fabrication of silica waveguides, but also for the straightforward application in quantum cryptography systems due to a long history of classical communications via low loss optical fibres \cite{Ten2016}. The downfall of working at telecom wavelengths resides in the use of high performance superconducting detectors working at cryogenic temperatures that limits the scalability and potential commercialization of these devices. Placing the sample in a cryostat also precludes the use of modulators based on the thermo-optical effects, commonly used in present silicon devices for quantum optics \cite{Silverstone2013}. In order to avoid working in cryogenic temperatures, we investigated photon pair generation in the near visible region. The possible integration of silicon avalanche photodetectors would result in a fully integrated quantum photonic chip working at room temperature.

Silicon nitride (SiN) is a CMOS compatible material \cite{Moss2013} with a high refractive index of 2 that exhibits moderate $\chi^{\left(3\right)}$ nonlinearity. It has been used both to generate single photons \cite{Ramelow2015} as well as to manipulate quantum states \cite{Xiong2015, Mohanty2017} at telecom wavelengths. Thanks to the high bandgap of $~5 eV$, SiN is the ideal candidate to design SFWM sources near the visible range and promises direct integration with silicon components. In this Letter we demonstrate generation rates of more than 2 million pairs of photons in the near visible range in a small ($19 \mu m$ radius), low loss ($2.2 dB/cm$) SiN ring resonator. The source developed here can find application in quantum networks, as the wavelengths and bandwidths are compatible with solid-state quantum memories \cite{Saglamyurek2011}, systems based on Raman schemes with hot atomic vapors \cite{Reim2010} as well as free space communication channels \cite{Yin2017}.

\begin{figure}[b]
 \centering
   \includegraphics[width=0.53\linewidth]{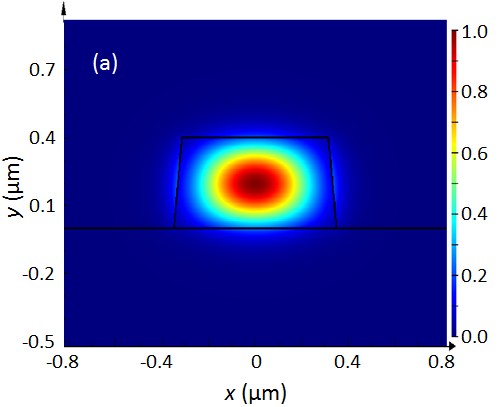}
   \includegraphics[width=0.43\linewidth]{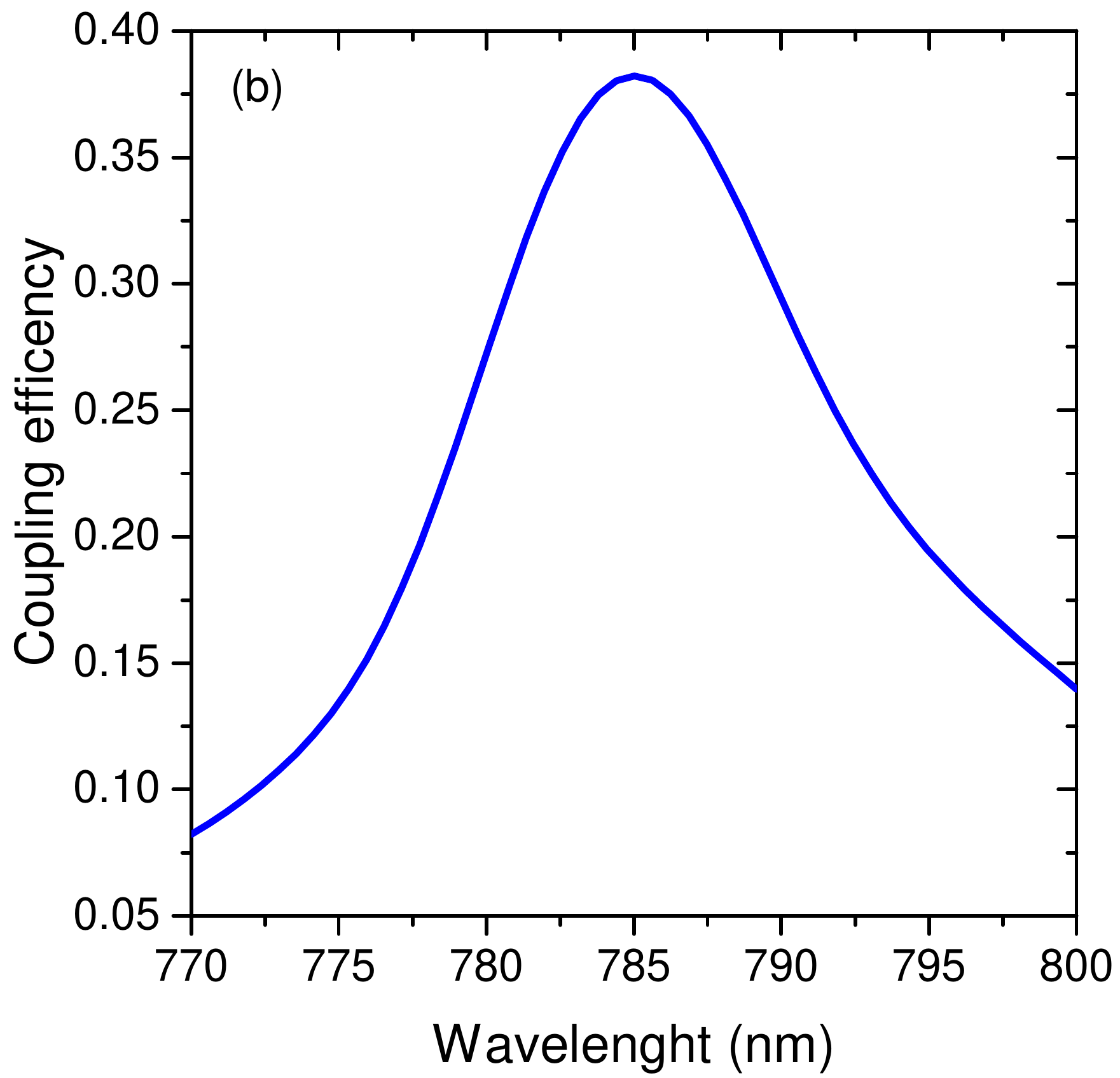}
\caption{(a) Fundamental TE mode of the SiN waveguide. (b) Coupling efficiency of apodized grating coupler.}
\label{Simulation}
\end{figure}

In order to minimize the field overlap with the waveguide sidewall and optimize confinement and loss propagation, we designed devices with dimensions of $400 x 700 nm$, showing multi-mode behavior. The fundamental TE mode used for this experiment is reported in (Fig. \ref{Simulation}(a)). To couple light from SiN waveguides to optical fibers, we optimized apodized gratings following the algorithm described in reference \cite{Martini2017}. In order to reduce lithography steps we designed the gratings to be fully etched, while apodization maintained a simulated coupling efficiency of 38\% and a $3 dB$ bandwidth of more than $20 nm$, as shown in Fig. \ref{Simulation}(b). Mode converters are used to reshape the optical mode adiabatically so that losses are minimized while changing the transverse size of the waveguide from $13 \mu m$ to $700 nm$ in a length of 350 $\mu m$.

\begin{figure}[tb]
 \centering 
  \includegraphics[width=0.48\linewidth]{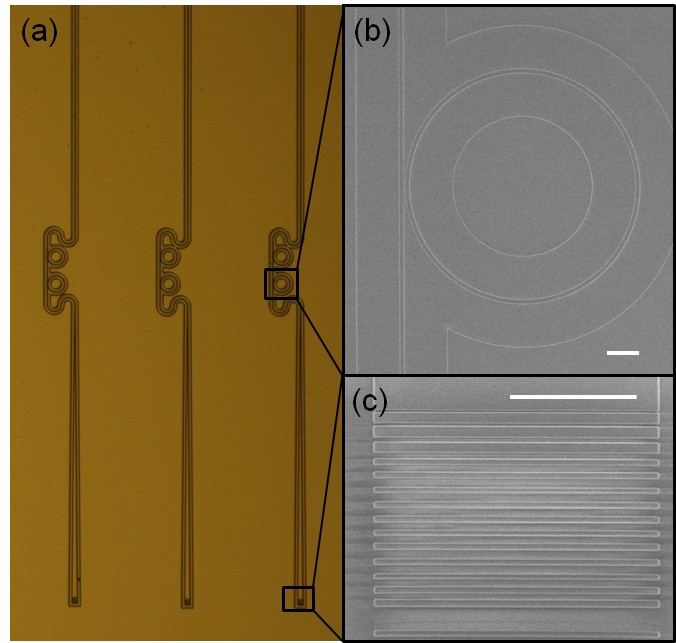}
   \includegraphics[width=0.48\linewidth]{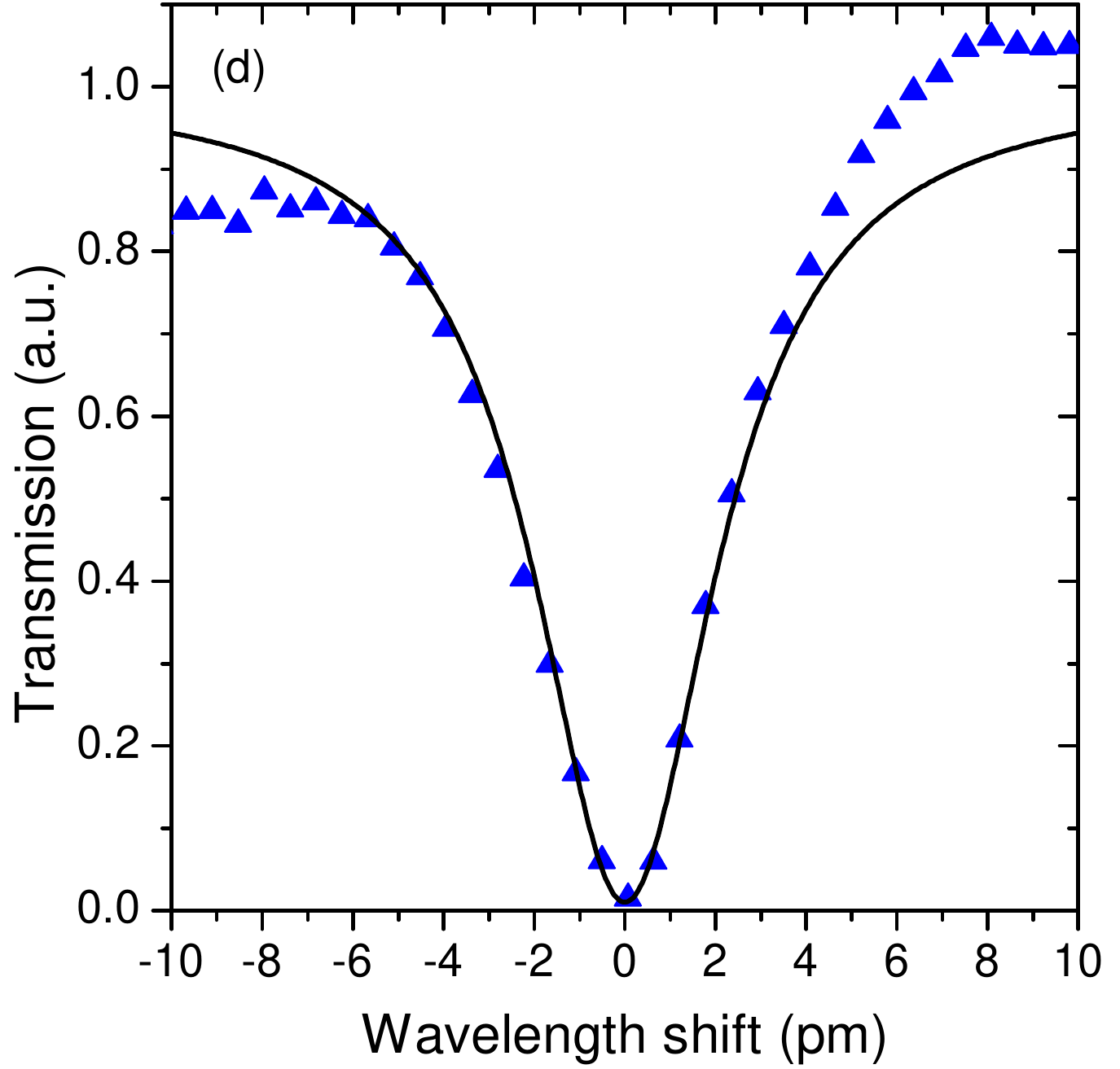}
\caption{(a) Optical picture of the fabricated photonic chip, with SEM pictures illustrating (b) the ring resonator and (c) apodized grating coupler (scale bars correspond to $5 \mu m$). (d) Transmission measurement of the ring resonator used for SFWM (dots), and best fit (solid line) showing an intrinsic Q-factor of 320,000.}
\label{sample}
\end{figure}

The photonic devices were fabricated starting from $2 \mu m$ of low loss thermal SiO$_{2}$ grown by wet oxidation on top of a standard silicon wafer. Then we deposited $410 nm$ PECVD SiN with a 5:2 ratio of NH$_{3}$:SiH$_{4}$ to minimize the amount of silicon nanoclusters \cite{Gorin2008}, showing a refractive index of 1.97 at $785 nm$. Samples were diced and spun with $450 nm$ of CSAR, before being exposed with an electron beam lithography system JEOL JBX-9300FS. After the lithography step we developed the resist and etched the sample with ICP RIE. We removed the resist leftover and deposited $1.2 \mu m$ of PECVD TEOS SiO$_{2}$ as top cladding. All the fabrication steps after the wet oxidation process were performed in temperatures not exceeding 350 degrees, making PECVD SiN compatible with Back End of Line CMOS processes. An optical image and SEM details of the photonic chip are reported in Fig. \ref{sample}(a)-(c). Devices were optically characterized by performing transmission measurements with a DFB tunable laser. Fig. \ref{sample}(d) displays the transmission of a critically coupled ring with $19 \mu m$ radius with an intrinsic Q-factor of 320,000. From this value we can extract the propagation losses of the structure to be $2.2 dB/cm$.

\begin{figure}[tbp]
\centering
\includegraphics[scale=0.29]{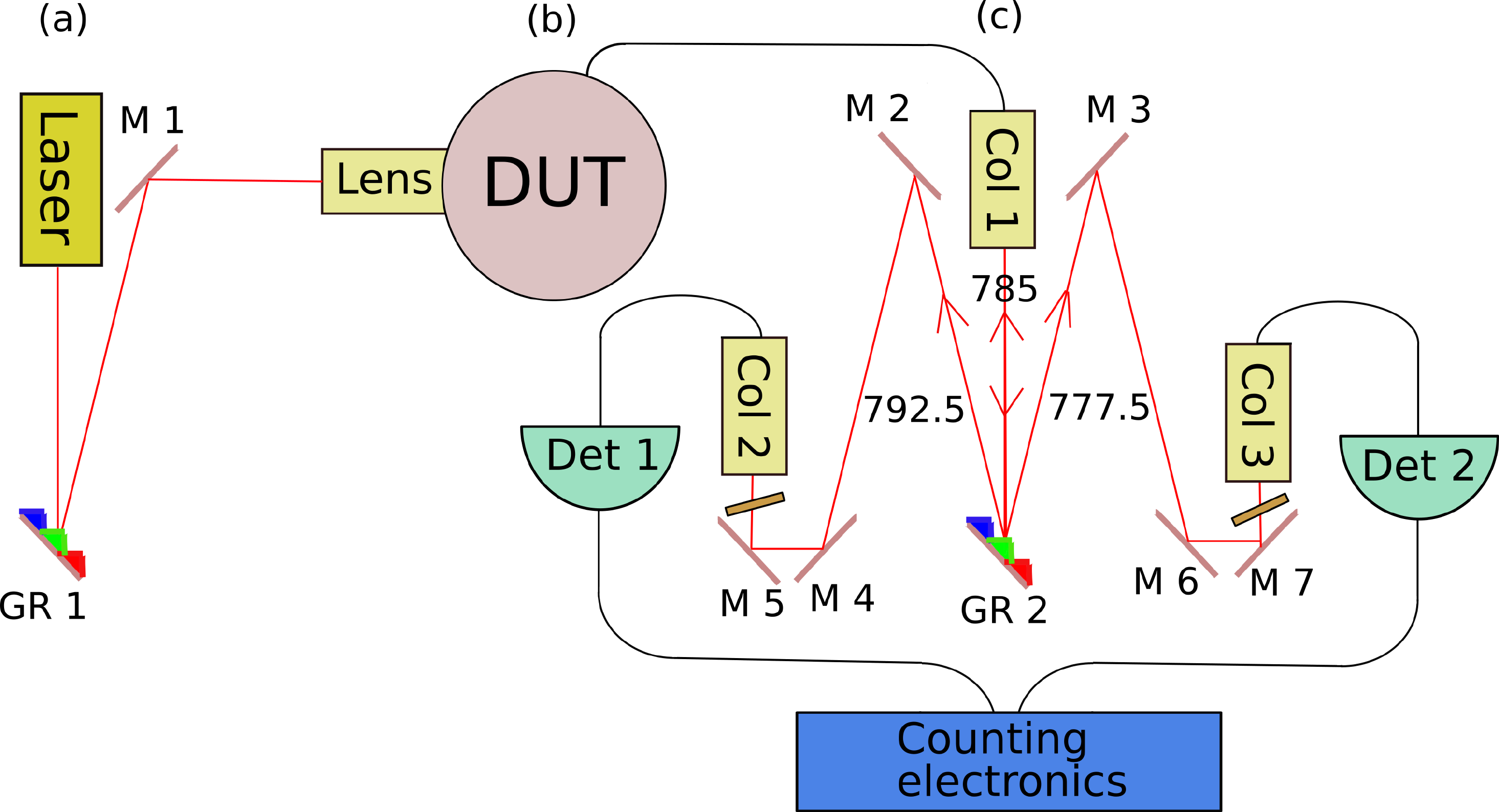}
\caption{Schematic setup for on chip generation and measurement of single photons. (a) Spectral filter of the noise generated from the pump laser. (b) Coupling to the SiN chip and generation of signal and idler photons. (c) Free-space optics to separate the generated photons and remove the pump stray light. GR: diffraction grating, M: mirror, Col: collimation optics.}
\label{setup}
\end{figure}

We designed an experimental setup to measure SFWM that consists of three parts (Fig. \ref{setup}). The first part in Fig. \ref{setup}(a) is composed by spectral filtering to reduce the laser sideband noise to the level of the single photons. 
This was achieved with one dispersive grating whose stray light reduction is $60 dB$ at frequencies of signal ($777.5 nm$) and idler ($792.5 nm$). 
The filtered beam was then coupled via a high NA objective to the $2 mm$ long input waveguide, shown in Fig. \ref{setup}(b). We used free space coupling for the input to avoid any spurious light generated in optical fibers. The pump laser was coupled to the ring resonator under test where it was attenuated by $23 dB$ due to the destructive interference of the resonator, and simultaneously the twin photons were generated. Signal, pump and idler photons were then coupled to a fiber array thanks to the grating coupler.
The last part of the setup, Fig. \ref{setup}(c), was used to separate the signal and idler photons and reject the pump light. In order to separate the correlated photons we used a dispersive grating in Littrow configuration where the pump wavelength is back reflected while the signal and idler are diffracted to opposite angles. Additional rejection was provided by 3 $nm$ bandwidth filters centered at respective signal/idler wavelength. The single photons were finally coupled to single mode fibers and detected by silicon avalanche photodiode detectors with quantum efficiency of 65\%.
The electric signals were recorded by counting electronics and a time interval analyzer to measure single and coincidence counts.

\begin{figure}[htbp]
\centering
\includegraphics[width=0.8\linewidth]{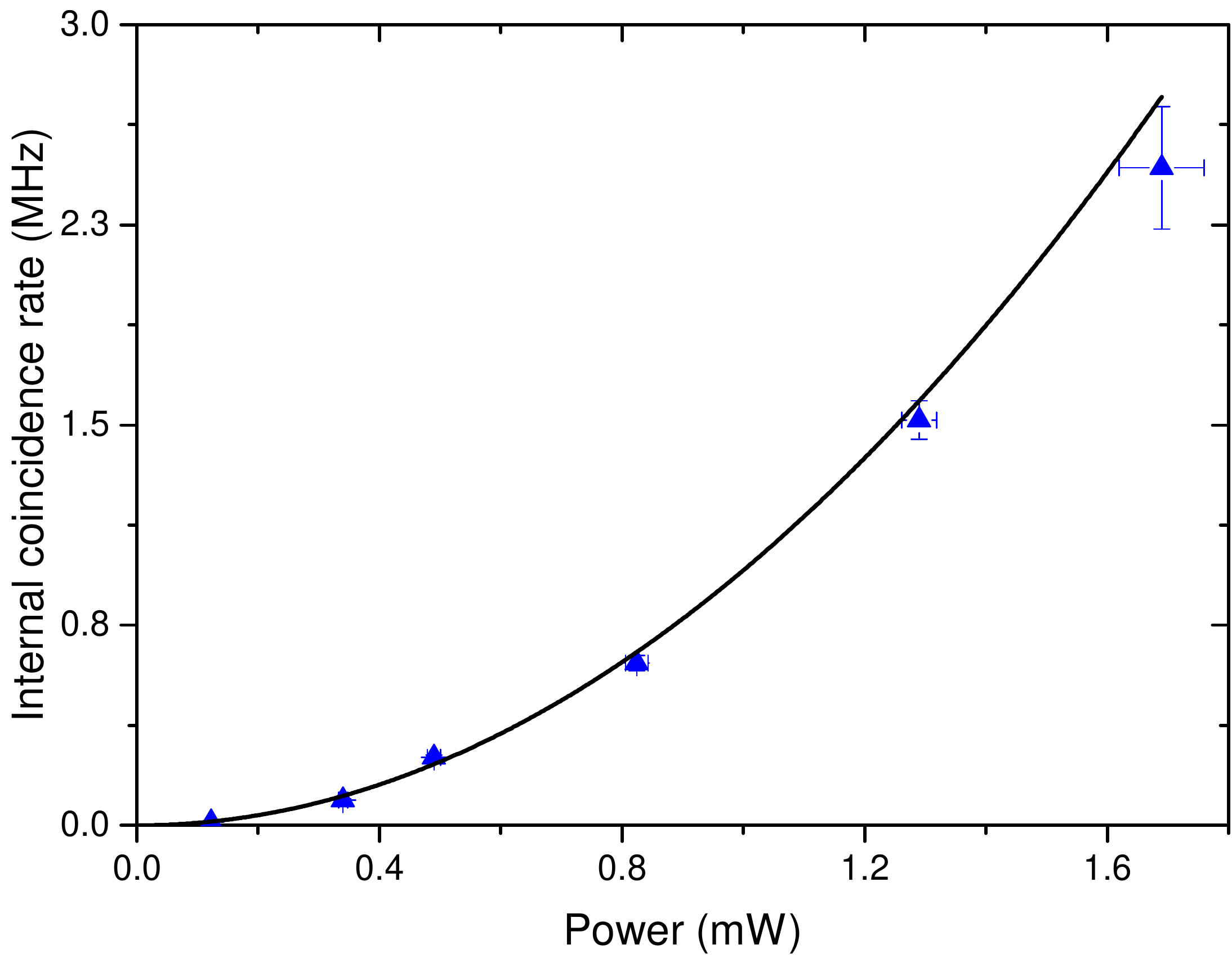}
\caption{Measured (dots) and theoretical (solid line) generation of photon pairs at the ring location as a function of the power coupled into the chip, measured with $1152 ps$ time integration.}
\label{SFWM}
\end{figure}

From the coincidence measurement we were able to extract the information about generated single photons in the ring coupled to the waveguide. The experimental data of generated pairs are extracted from the equation
\begin{equation}
G_{SFWM}=\frac{P^{SFWM}}{2\hbar \omega_{p}} = \frac{CC}{10^{-\left(\eta_{s}+\eta_{i}\right)/10}},
\label{rate}
\end{equation}
where $CC$ are measured coincidences, $\eta_{s/i}$ are the overall collection losses for the signal and idler, which in our case are $16.4 dB$
and $24.1 dB$, respectively. $P^{SFWM}$ is the power of measured photons based on the equation \cite{Helt2012}
\begin{equation}
P^{SFWM} = \left(\gamma L\right)^{2} \left(\frac{Q_{l}v_{p}}{\omega_{p}L/2}\right)^{3}\frac{\hbar\omega_{p}v_{p}}{2L}P_{p}^{2} ,
\label{eqSFWM}
\end{equation}
where $\gamma$ is the nonlinear parameter, $L$ is the circumference of the ring, $P_{p}$ is power of the pump in the ring, $Q_{l}$ is the measured loaded Q factor. The nonlinear parameter is defined in as the usual $\gamma = \frac{2\pi n_{nl}}{\lambda_{p}A_{eff}}$, where we used the nonlinear index $n_{nl}$ from literature \cite{Lacava2017}. The mode effective area $A_{eff}$ and group velocity $v_{p}$ of the fundamental mode were numerically calculated. We can conclude that the experimental data in Fig. \ref{SFWM} are in a good agreement with the theoretical model of Eq. \ref{rate}. This confirms the nonlinear relationship between the generated single photons and pump photons with a generation rate of $950,000 \frac{pairs}{mW}$.
 
\begin{figure}[htbp]
\centering
\includegraphics[width=\linewidth]{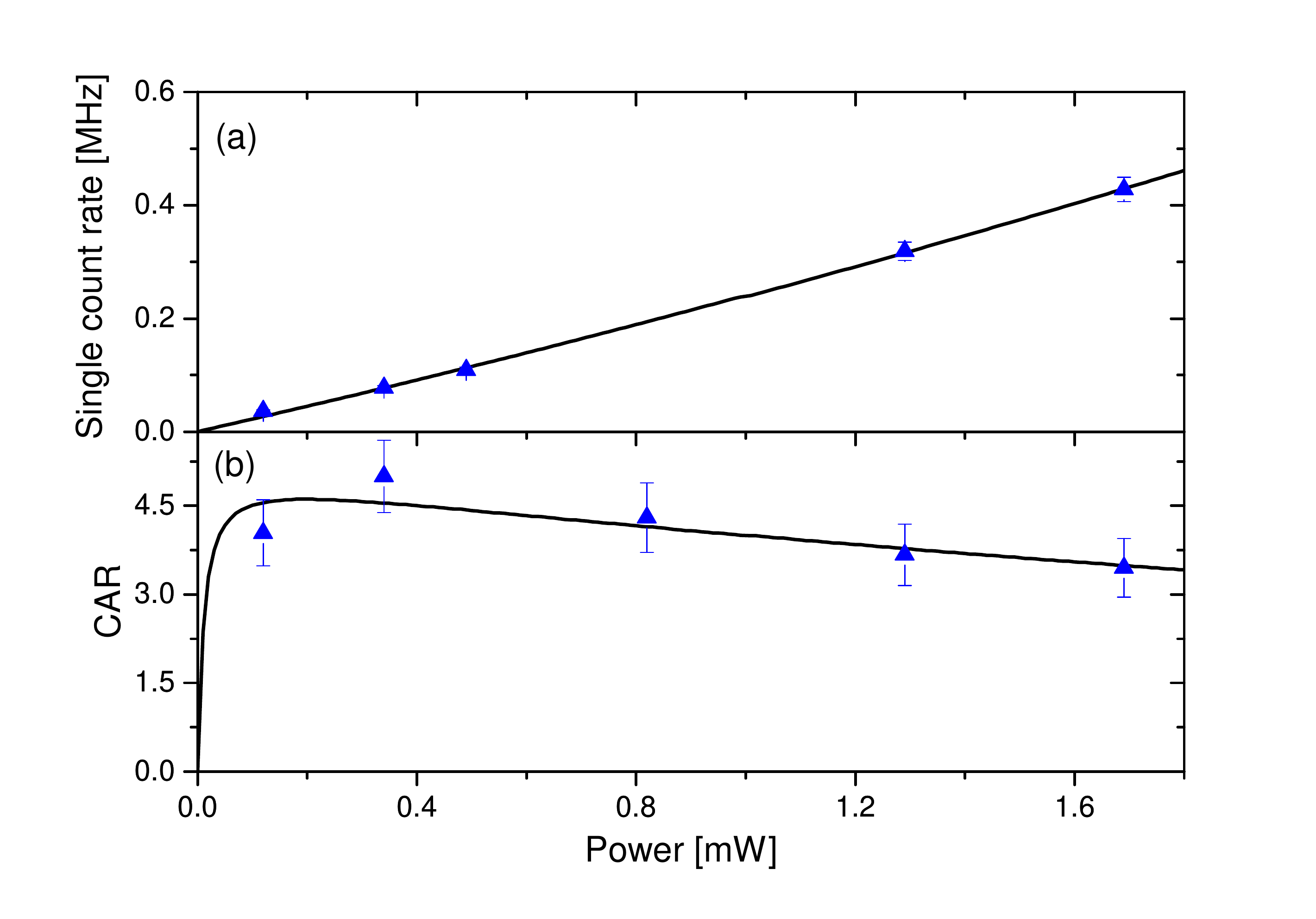}
\caption{(a) Measured (dots) and theoretical (solid line) power dependence of the single photon generation at the signal
wavelength (777.5nm) fitted by linear function. (b) Measured (dots) and theoretical (solid line) power dependence of the Coincidence to Accidental Ratio ($CAR$).}
\label{raman}
\end{figure}

Beside the photon pair generation rate, an important figure of merit to characterize the quality of a correlated photon pair source is the coincidence $CC$ to accidental $AC$ ratio $CAR$. It takes into account the relationship between the number of SFWM  photon pairs and the number of accidental coincidences, coming from photons emitted randomly in time 
\begin{align}\label{CAR}
CAR &=\frac{CC}{AC}, &&\\
CC &=G_{SFWM}\eta_{s}\eta_{i}, &&\\
AC &=R_{s}R_{i}\delta t, &&\\
R_{s} &=\left(n_{s}+G_{SFWM}\right)\eta_{s}+dc_{s}, &&\\
R_{i} &=\left(n_{i}+G_{SFWM}\right)\eta_{i}+dc_{i},
\label{CAR1}
\end{align}
where $CAR$ depends on the integration time $\delta t$ and dark counts of the detectors $dc_{s/i}$, and any linear noise $n_{s/i}$ resulting from the light interaction in the material. $R_{s/i}$ are the measured singles at signal and idler frequency, respectively. In Fig. \ref{raman}(a) we report the power dependence of the detected signal photons $R_{s}$, showing a linear dependence. This relation suggests that the ring resonator emits uncorrelated photons generated from spontaneous Raman scattering, as suggested by the  broad scattering spectrum of SiN \cite{Dhakal16}.

In Fig. \ref{raman}(b) we characterized the $CAR$ as a function of the power in the waveguide. The number of coincidences and accidental counts is measured with a time window of 1152 $ps$, longer than the FWHM value of the coincidence peak of $\sim 500 ps$. This is to ensure that all the coincidence counts are included, at the expense of a lower $CAR$ \cite{Engin2013}. Moreover, the FWHM value is mainly dictated by the jitter of our Si detectors, rather than the lifetime of the photons ($< 200 ps$, considering the Q factor of our cavity). Faster detectors, or pulsed excitation, would result in a higher measured $CAR$.
Fig. \ref{raman}(b) also includes a curve from equations \ref{CAR}-\ref{CAR1} where the parameters are extracted from Fig. \ref{SFWM}, considering correction of signal and idler photons by respective phonon distribution \cite{Takesue2005},  and \ref{raman}(a) without fitting any additional variable. The model shows good agreement with data, indicating that the $CAR$ value is limited from Raman photons at all the powers used in the experiment. Effects from detector dark counts are only visible at very low pump powers, while multi-photon events cause a small decrease in $CAR$ for increasing power in the tested range. These results are comparable to CAR measurements in strip SiN waveguides \cite{Zhang16} and Hydex rings resonators \cite{Reimer14} at telecom wavelengths, once the detection timing is considered.

In conclusion, we studied spontaneous four wave mixing process in the near visible region generated in a microring resonator. Contrary to alternative sources based on periodically pooled crystals \cite{Sanaka01, Banaszek01} the CMOS compatible fabrication process would make possible the integration of non-classical sources together with silicon based detectors working at room temperature.
We believe that the measured $CAR$ is mainly limited by the linear noise coming from uncorrelated Raman photons which affects amorphous materials, including Silicon Nitride. Optimization of the optical properties of SiN through tailored material deposition, combined with the design parameters for the ring resonator, can lead to an increase of generation rate and $CAR$ value. An increase of Q factor will also reduce the bandwidth of the generated photons from the current $\sim 2.4 GHz$ to sub-$GHz$ values, making this source fully compatible with atomic memories. Finally, the confirmed absence of saturation in the pair generation rate, thanks to the high bandgap of SiN, suggests that integrated single photon sources can be developed well in the visible spectrum. These properties make SiN resonators excellent candidates to expand the applications of integrated single photon sources for a broad range of quantum technology applications.

\section*{Funding Information}
This work was supported by the H2020-FETPROACT-2014 Grant QUCHIP (Quantum Simulation on a Photonic Chip; grant agreement no. 641039).

\section*{Acknowledgments}
We acknowledge support from the Southampton Nanofabrication Centre.


\end{document}